\documentclass[usenatbib,letters,fleqn]{mn2e}
\usepackage{amsmath,amssymb}
\usepackage{graphicx}
\usepackage{epstopdf}
\graphicspath{{./figures/}}
\usepackage{url}
\usepackage{aas_macros}
\usepackage{astro}
\usepackage[subtle]{savetrees}

\newcommand{\eddr}{\dot{M}/\dot{M}_{\rm Edd}}
\newcommand\lsim{\mathrel{\rlap{\lower4pt\hbox{\hskip1pt$\sim$}}
        \raise1pt\hbox{$<$}}}
\newcommand\gsim{\mathrel{\rlap{\lower4pt\hbox{\hskip1pt$\sim$}}
        \raise1pt\hbox{$>$}}}

\author[Generozov \& Haiman]{Aleksey Generozov\thanks{E-mail: (alekseygenerozov,zoltan)@astro.columbia.edu} and 
Zolt\'an Haiman$\footnotemark[1]$\\
Department of Astronomy, Columbia University, 550 West 120th Street, New York, NY 10027}

\begin{document}
\title{Lyman edges in supermassive black hole binaries}
 \maketitle

\begin{abstract}
We propose a new spectral signature for supermassive black hole
binaries (SMBHBs) with circumbinary gas discs: a sharp drop in flux
bluewards of the Lyman limit.  A prominent edge is produced if the gas
dominating the emission in the Lyman continuum region of the spectrum
is sufficiently cold ($T\lsim20,000$K) to contain significant neutral
hydrogen. Circumbinary discs may be in this regime if the binary
torques open a central cavity in the disc and clear most of the hot
gas from the inner region, and if any residual UV emission from the
individual BHs is either dim or intermittent.  We model the vertical
structure and spectra of circumbinary discs using the radiative
transfer code \texttt{TLUSTY}, and identify the range of BH masses and
binary separations producing a Lyman edge.  We find that compact
supermassive ($M\gsim 10^{8}~{\rm M_\odot}$) binaries with orbital
periods of $\sim 0.1 - 10 \yr$, whose gravitational waves are
expected to be detectable by pulsar timing arrays, could have
prominent Lyman edges.  Such strong spectral edge features are not
typically present in AGN spectra and could serve as corroborating
evidence for the presence of an SMBHB.
\end{abstract}


\section{Introduction}
\label{sec:introduction}

Supermassive black holes (SMBHs) are present in the centres of most,
if not all, nearby galaxies (see reviews by e.g. \citealt{KR95};
\citealt{FF05}).  If two galaxies containing SMBHs merge, this should then result in the formation of a
SMBHB (e.g. \citealt{Begel:Blan:Rees:1980}). Thus, given the
hierarchical model for structure formation, in which galaxies are
built up by mergers, one would naively expect SMBHBs to be
quite common.

Many candidates for binary BHs have been
identified on kiloparsec scales, including two galaxies with spatially
resolved active binary nuclei (\citealt{Komossa+2003,Fabbiano+2011};
see, e.g. the review by \citealt{Komossa:Rev06} and \citealt{Shen+:2013} and references therein).  At parsec scales,
however, there is only one clear example: a radio observation of a BH pair with
a projected separation of $\sim$7 pc \citep{Rodriguez+:2006}. There
remain no confirmed binary black holes at subparsec separations.

The lack of observational evidence for binaries at small separations
suggests that the SMBHs either remain inactive during the merger, or
that they merge within a small fraction of a Hubble time and are
consequently rare (\citealt{HKM09}).  Another possibility is that the
spectrum of a compact binary differs significantly from those of
single-BH active galactic nuclei (AGN). A better understanding of the
spectral energy distributions (SEDs) and lightcurves from circumbinary
discs is necessary to determine whether binaries may therefore be
missing from AGN surveys or catalogs \citep{Tanaka:2013}.

Gravitational waves (GWs) from a merging SMBHB may be
detected in the next decade by pulsar timing arrays (PTAs;
\citealt{PTAs}).  Identifying the gravitational wave source in 
EM bands would also have considerable payoffs for cosmology and 
astrophysics (e.g. \citealt{PhinneyWhitePaper}). Unfortunately, GWs yield limited precision on the sky position. For a PTA source, of
order 10$^2$ (and perhaps as many as 10$^4$) plausible candidates may
be present within the 3D measurement error box
\citep{Tanaka:2012}. Concurrent EM observations would then be
necessary to identify the GW source.

Many different EM signatures have been proposed for SMBHBs.
These include periodic luminosity variations commensurate with the
orbital frequency of the source (\citealt{Haiman+2009} and references
therein) and broad emission lines that are double-peaked and/or offset
in frequency (\citealt{Shen+:2013} and the references therein).
Additionally, the evacuation of a central cavity by the binary could
lead to a spectrum that is a distinctively soft
\citep{Milos:Phinney:2005,TanakaMenou:2010}, and has unusually weak
broad optical emission lines, compared to typical AGN
\citep{Tanaka:2012}. Other work \citep{Gultekin:Miller:2012, Roedig+:2014}
describes spectral signatures of discs with partial cavities, which may show up in 
the SED as broad, shallow dips.

We here propose that, in addition to the above signatures, the presence
of a central cavity in a circumbinary disc could produce distinct
absorption edges in the optical/UV (in particular, at the Lyman
limit). This is analogous to the prominent Lyman break in galactic
spectra a $\gsim$ few$\times10$ Myr after a starburst, when the
composite emission is dominated by less massive stars with cooler
atmospheres \citep{Leitherer+1999,Schaerer2003}.  Physically, such an
edge would be present if the disc is cold enough
($T_{\eff}\lsim20,000$ K) to have sufficient neutral (ground-state)
hydrogen to absorb Lyman continuum photons. The disc should also be
hot enough (conservatively $T_{\eff}\gsim10,000$ K); for yet cooler
discs, the continuum emission redward of the Lyman limit may be
obscured by metal absorption features.

This Letter is organized as follows.
In\,\S\ref{sec:discmodels}, we describe the details of our disc and
emission models.
In\,\S\ref{sec:results}, we show examples of spectra for binary BH
discs, and compare these to those of single BH discs.
In\,\S\ref{sec:minidiscs}, we discuss caveats, including emission from
mini-discs around each of the individual BHs that could mask the Lyman
edge.
We summarize our main conclusions and the implications of this work in\,\S\ref{sec:conclusion}.

\vspace{-\baselineskip}
\section{Disk Models}
\label{sec:discmodels}

In order to model disc spectra, we must begin with a model for the
disc. In particular, we need the energetics (i.e. how much
energy is dissipated from tidal torques and viscous heating throughout
the disc).  We here adopt the analytic models in \citet[][hereafter
KHL12]{Kocsis+2012a}.  These are modified versions of a standard
\citet{SS73} accretion disc, to incorporate the angular momentum
transfer and the corresponding heating of the disc by the binary
torques. The models self-consistently track the co-evolution of the
disc and the binary orbit, through a series of quasi-steady,
axisymmetric configurations.  

There are several qualitatively different solutions for such a system,
depending on the parameters of the binary (i.e. masses and
orbital separation) and the disc (i.e. viscosity and accretion
rate).  For the purposes of this Letter, we concentrate on
the case where a central cavity is opened and
maintained, as the lack of the hot inner regions is responsible for
the Lyman edges in the spectrum. Our conclusions on Lyman edges simply rely on emission from disc patches with effective temperature $T_{\rm eff}\approx
10,000-20,000$K dominating the composite UV spectrum, and should not be sensitive to model details.

\begin{figure}
 \centering \mbox{\includegraphics[width=\linewidth]{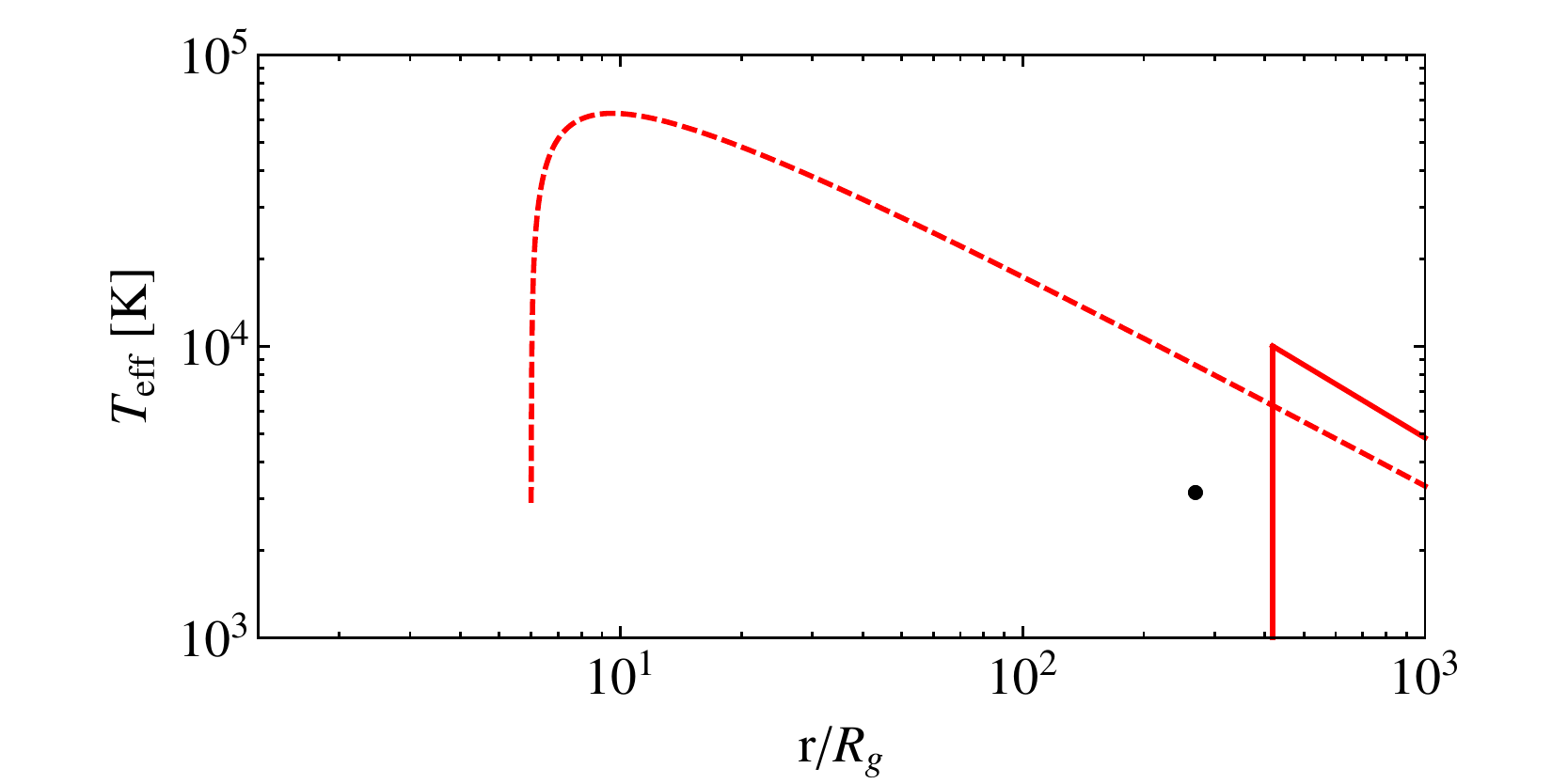}}
 \caption{Temperature profiles for a circumbinary disc around a 10$^8
   \Msun,\; q=0.05$ binary (solid) and for a thin disc around a single
   10$^8 \Msun$ BH (dashed). Both discs assume an accretion
   rate $\dot{M}/\dot{M}_{\rm Edd}=0.25$, and the radial distance is
   shown in gravitational units.  The solid curve is truncated at a
   radius of 470 $R_g$ due to the presence of a cavity in the disc. The black 
   dot marks the radius of the secondary.}
 \label{fig:temperature}
\end{figure}

Fig.~\ref{fig:temperature} shows illustrative examples of radial
profiles of the effective temperature (defined by $\sigma_{\rm SB}
T^4_{\eff}\equiv F$, where $\sigma_{\rm SB}$ is the Stefan-Boltzmann
constant, and $F$ is the total dissipation rate per unit disc surface
area, including both viscous and tidal heating).  The solid curve
shows $T_{\rm eff}(r)$ for a circumbinary disc around a $M_{\rm
  tot}=M_1+M_2=10^8 \Msun$ binary, with mass ratio $q=M_2/M_1=0.05$
and accretion rate $\eddr=0.25$ (assuming a
radiative efficiency of 10\%).  The secondary is located at the radius
$r_s =$ 230 $R_g $ ($R_g \equiv G M_{\rm tot}/c^2$), creating a cavity
inside 470 $R_g$.  We assume in all of our models that the BHs
are non-spinning, and adopt a viscosity parameter $\alpha=0.1$.
For the other less important parameters, we use the same fiducial parameters as KHL12, 
except we set $f_{\rm T}=3/8$ to be consistent with our assumption of vertically uniform dissipation
The dashed curve shows, for comparison, $T_{\rm eff}(r)$ for a disc
around a single BH with the same mass.  Most importantly, this figure
shows that outside the cavity, the circumbinary disc is hotter (by a
factor of $\sim$two; see also \citealt{Lodato+09}) than the
corresponding single-BH disc, but still not nearly as hot as the
innermost regions of this single-BH disc.

\subsection{SEDs: previous models}

In the simplest model for the emerging spectrum, each disc annulus emits as a blackbody,
with the effective temperature determined by the heating rate from
viscous and tidal torques in that annulus.  A more sophisticated
version is a composite ``greybody'' spectrum. The blackbody emission
in each disc annulus is modified by a correction factor, accounting
for the effects of electron scattering opacity
\citep{Milos:Phinney:2005,TanakaMenou:2010}, to obtain
\begin{equation}
 F_\nu \sim \pi \frac{2 \epsilon_{\nu}^{1/2}}{1+\epsilon_{\nu}^{1/2}} B_{\nu}.
\end{equation}
Here, $\epsilon_{\nu} \equiv \kappa_{\rm abs, \nu}/\left(\kappa_{\rm
  abs, \nu}+\kappa_{\rm es, \nu}\right)$ is the ratio of absorption to
total opacity. For this model, we adopt the Kramers' bound-free opacity at solar
metallicity for $\kappa_{\rm abs, \nu}$ and $\kappa_{\rm es, \nu}=0.4$
cm$^{2}$ g$^{-1}$.  We refer the reader to
\citet{TanakaMenou:2010} for a detailed discussion of the greybody
disc model.

\subsection{SEDs: new RT models}

The greybody model ignores important radiative transfer (RT) effects. In particular, the
opacity is assumed to be a smooth function of frequency. In reality,
the bound-free opacities have sharp thresholds, corresponding to the
onset of absorption from various species in the disc.  For example,
photons with energies above the Lyman limit $\geq$ 13.6eV (or
frequency 3.28 $\times$ 10$^{15}$ Hz) can be absorbed by neutral
(ground-state) hydrogen (H) within the disc.  This sharp change in
opacity can cause a corresponding prominent edge in the emerging disc
spectrum. This may be understood as follows: in a plane-parallel infinite atmosphere, one sees the
source function at optical depth $\sim$unity along the line of
sight. Thus, the smaller opacity redwards of the Lyman limit means that we
see the source function from deeper in the disc, where the temperature
is (generally) higher. The higher temperature then translates into a
greater flux redward of the Lyman limit 
(e.g. \citealt{HubenyHubeny:97}).

As the temperature increases, H is more ionized, decreasing the H
bound-free opacity.  Eventually, the combined electron scattering and
free-free opacities overwhelm the discontinuity in the bound-free
opacity, and the opacity on both sides of the Lyman edge becomes
nearly equal, washing out any absorption edge in the spectrum. As the
temperature continues to increase, non-local thermodybnamic equilibrium (NLTE) effects may cause an emission
edge instead.

To model the emission from the disc, we use the RT
code \texttt{TLUSTY} (\citealt{HubenyLanz:95}). This code
self-consistently solves the equations of vertical hydrostatic
equilibrium, energy balance, RT, and the full non-LTE
statistical equilibrium equations for all species that are present in
the disc. Contributions from all bound-free and free-free transitions
at all frequencies of interest are included, while bound-bound
transitions are assumed to be in detailed balance.

We model H as a nine-level atom and He as a four-level atom.
Electron scattering, including Comptonization is also included. We
assume that the disc is composed of H and He (at their Solar ratio). Metals
are not included, but would make little difference to continuum
spectra for $10^4$ K $\lesssim \teff \lesssim 10^5$ K.  For any given
annulus, we calculate the vertical structure by specifying the
vertical gravity $g_z$, disc surface density $\Sigma$, and the total
energy dissipation rate $\teff$.  Spectra are insensitive to the
surface density (provided the disc remains optically thick) and, for
computational convenience, we fix $\Sigma=2\times10^5~{\rm g~cm^{-2}}$
throughout this {\it Letter}. Thus, specifying the radial profile
$\teff(r)$ and vertical gravity $g_z(r)$ fully determines the spectrum
emerging from the disc annulus at radius $r$. $g_z(r)$ is approximated to be linear in
$z$ and proportional to the square of the local Keplerian angular
frequency, $q_{g}\equiv\Omega_{\rm K}^2$.  The composite spectrum
may be computed by summing over all annuli. 

Although we have assumed that flux is radiatively transmitted through the
disc, most of our model annuli have convectively unstable zones.
Models with small density inversions (density increasing outward) also occur. These would be unstable. Finally, we
assume that tidal and viscous torques dissipate energy locally, and
uniformly in height (i.e. equal dissipation per unit column mass).
The vertical energy distribution is poorly understood, and real discs
may be advective, or the energy may be carried away by density waves
\citep{DRS2011,DM2012}.

\section{Results: Edges in Disk Spectra}
\label{sec:results}

In this section, we show examples of composite circumbinary disc
spectra, and compare these to the simpler black-- and greybody models,
as well as to the corresponding single--BH disc spectra.

\begin{figure}
 \centering
 \mbox{\includegraphics[width=0.96\linewidth]{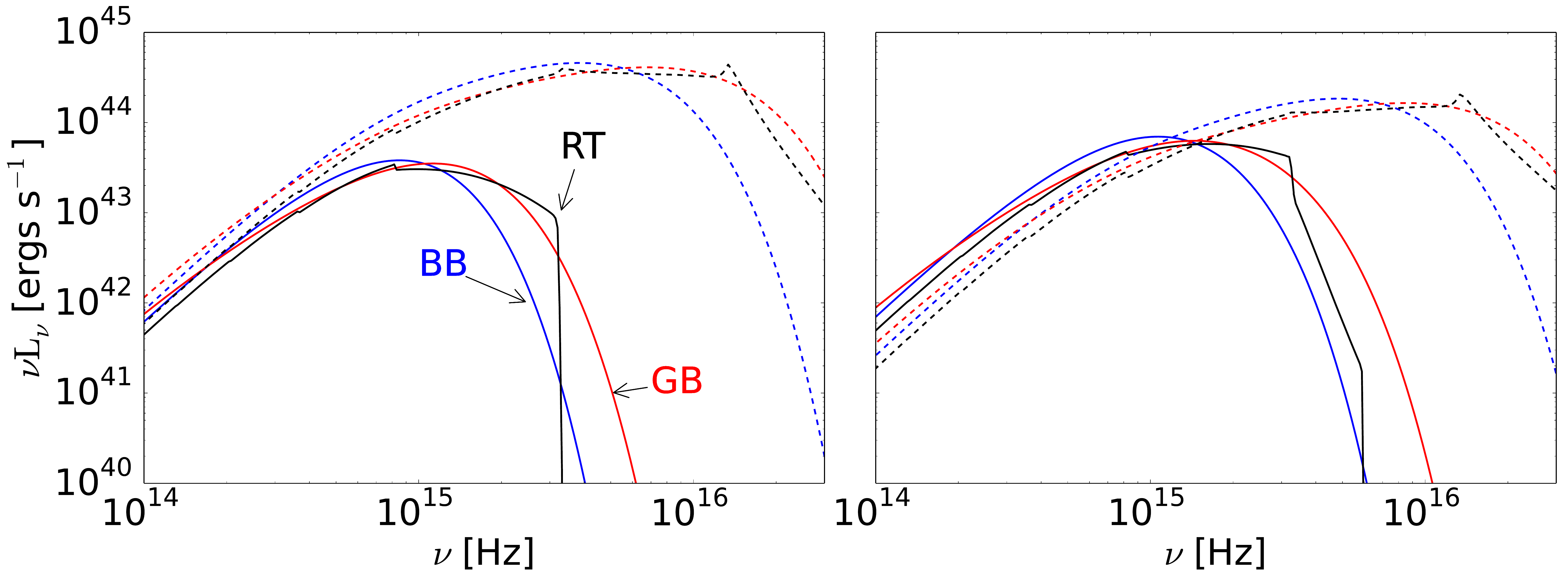}}
 \caption[]{{\it Left-hand panel:} Spectra from discs with the temperature
   profiles shown in Figure~\ref{fig:temperature}. The black, blue and
   red solid curves are \texttt{TLUSTY}, blackbody, and greybody
   spectra for circumbinary discs. The dashed curves with the same
   colors refer to the same models for a single-BH disc. {\it Right
     panel:} Same as the left, except for discs around 2.5 times
   lower-mass ($M_{\rm tot}=4 \times 10^7 \Msun$) BHs.}
 \label{fig:composite}
\end{figure}

In the left-hand panel of Fig.~\ref{fig:composite}, the solid (dashed)
curves show composite disc spectra corresponding to the circumbinary
(single-BH) disc in Fig.~\ref{fig:temperature}.  Realistically, the
disc may extend to 2000 $R_g$, where it would
become gravitationally unstable, according to the Toomre criterion.
However, due to practical issues with convergence, we were only able
to get models within $\sim$40 $R_g$ of the inner disc edge, and we
only integrate the emission from this region.  However, the excluded
region is cooler than the inner edge of the disc and its emission
should not mask the Lyman edge.

The \texttt{TLUSTY} binary spectrum (in black), shows a prominent
break at the Lyman limit ($3.28\times10^{15}$ Hz), which is not
present in the black-- or greybody disc models (shown in blue and red,
respectively).  Overall, the blackbody model overpredicts the flux at
both low and high frequencies, but underpredicts it immediately below
the Lyman limit. The greybody model does a better job at frequencies
below the Lyman limit, but overpredicts the flux more significantly at
higher frequencies.

In contrast to the binary-BH spectrum, the single-BH spectrum has no
sharp absorption edge at the Lyman limit (although there is a weak
kink). This is because, as shown in Fig.~\ref{fig:temperature}, the
innermost region of the single-BH disc is much hotter than the inner
edge of the binary disc. Specifically, the binary disc has a maximum
$T_{\eff}$ of 11,000, whereas, the single-BH disc is considerably hotter,
with a maximum $T_{\eff}$ of 80,000 K (we have chosen a somewhat high 
$\eddr$ so that inner edge of the circumbinary disc
is above $10^4 K$. However, we have verified that there is no Lyman edge in the 
single BH case for $\eddr\sim0.1$ ).  For a spinning BH, the
inner edge is expected to be closer to the BH horizon
(following the innermost stable circular orbit), which would further
increase the maximum temperature in the single-BH case. However, winds may limit
the maximum $T_{\eff}$ in single-BH discs to 50,000 K~\citep{LaorDavis:2013a}.

In the right-hand panel of Fig.~\ref{fig:composite}, we show another set of
illustrative disc spectra, with the same parameters as in the left
panel, except for a lower mass ($M_{\rm tot}=4\times10^7\Msun$).  In
this case, the binary disc is hotter ($\teff=17,000$K at the inner
edge) and the Lyman edge feature is correspondingly weaker. The single-BH model
looks similar to the one shown in the left.

\subsection{Binary parameter space with edges}
\label{subsec:param_space}

We now discuss under what conditions binary discs are likely to have
prominent Lyman edges.

If the inner edge of the disc falls below a particular
threshold temperature, the composite disc spectrum will have a
prominent Lyman absorption edge. An example of a spectrum close to the
threshold temperature is the one shown in the right-hand panel of
Figure~\ref{fig:composite}: if the disc were much hotter, the Lyman
edge feature would be wiped out. The precise threshold will depend on
the vertical gravity and other parameters. Physically, the vertical
gravity sets the scale height, which determines the vertical density. The
density, in turn, determines the ionization state, which then affects
the opacity difference between the two sides of the Lyman limit.

To establish a quantitative criterion for the threshold
$\teff$, we find the maximum effective temperature such that there is at least
an order of magnitude drop in the flux at the Lyman limit (for different gravity parameters).  We then
establish the following fitting formula between the threshold
temperature as a function of the square of the local Keplerian angular
frequency, $q_{g}\equiv\Omega_K^2$:
\begin{equation}
\begin{split}
\log\left(\frac{T_{\rm thres}}{17,000\text{K}}\right)&=
0.06~\log\left(\frac{q_g}{10^{-12} s^{-2}}\right).
\label{e:hithres}
\end{split}
\end{equation}

\begin{figure}
 \centering 
 \mbox{\includegraphics[width=0.48\linewidth]{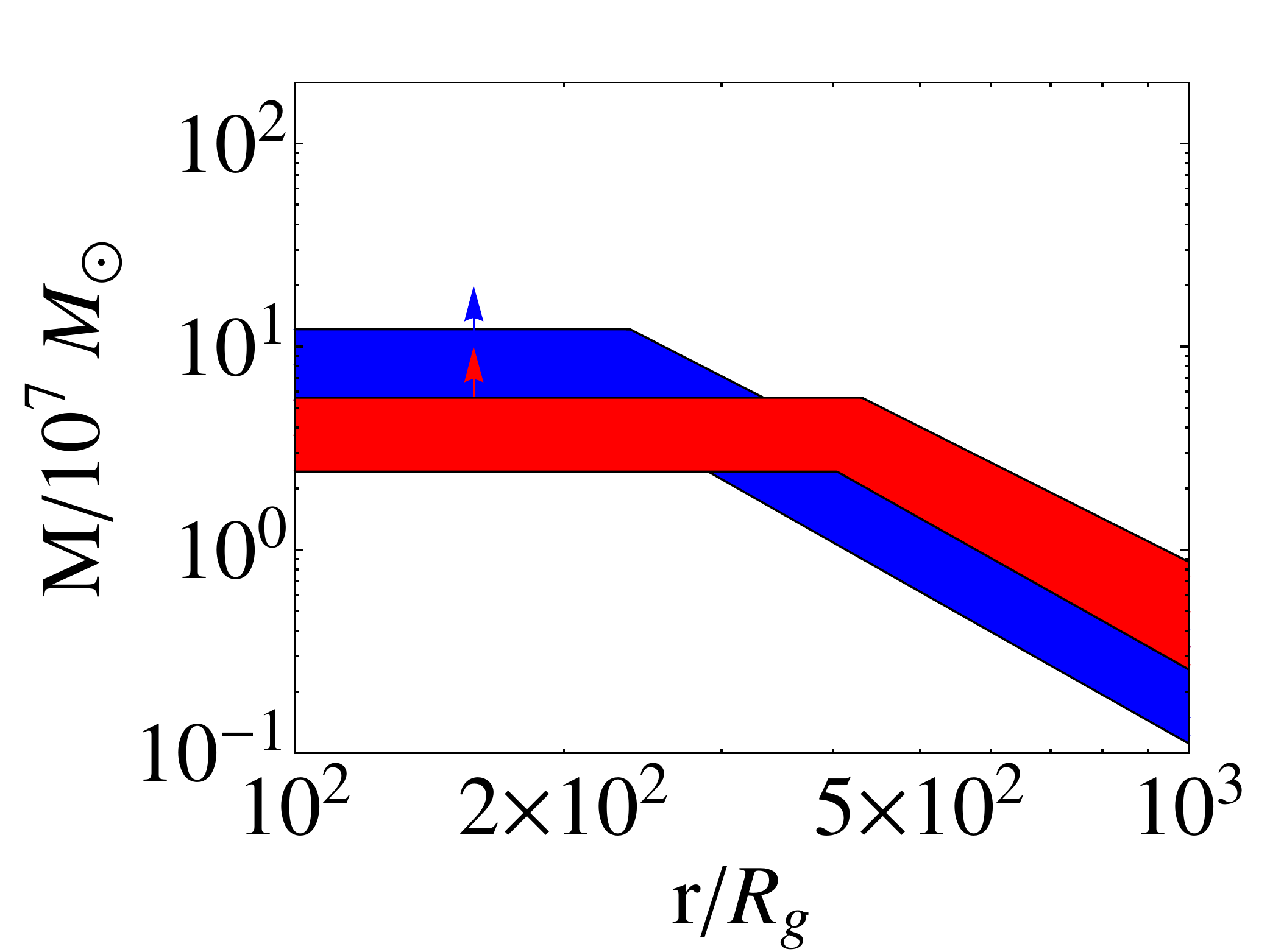}}
 \mbox{\includegraphics[width=0.48\linewidth]{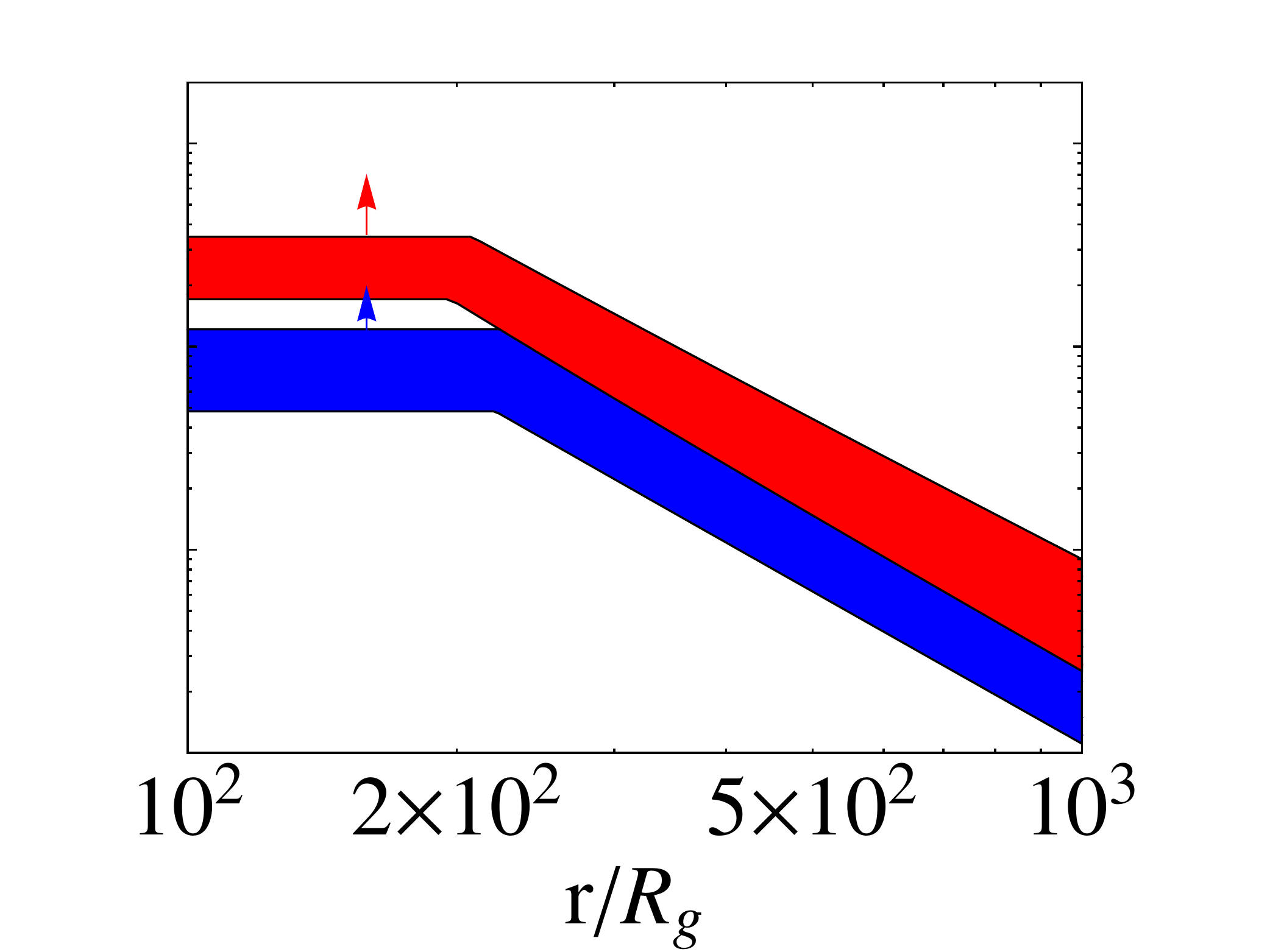}}
 \caption{Regions of binary parameter space with a prominent Lyman
   edge. In both panels, the blue region corresponds to binaries with
   a prominent Lyman edge for $q=0.05$ and $\eddr=0.25$. On the left,
   the red region corresponds to Lyman edges for $q=1$ and
   $\eddr=0.25$. On the right, the red region corresponds to
   $q=0.05$ and $\eddr=1$.  The bends below
   $r\lsim{\rm few}\times100 R_g$ correspond to binary separations
   where the disc structure has decoupled from the GW-driven
   binary. Above each colored region the inner disc temperature falls
   below $10^4$ K. Below $10^4$ K metal absorption could mask the Lyman edge, as indicated by the arrows.}
 \label{fig:param_space}

\end{figure}

This is accurate to $\sim1\%$\footnote{Assuming $\Sigma=2\times 10^5$ g cm$^{-2}$.
In the optically thick limit changes in the $\Sigma$ may affect the threshold by $\sim5\%$} in the range $10^{-12}
s^{-2}<q_g<10^{-9} s^{-2}$. At higher $q_g$, the threshold temperature
increases less steeply with $q_g$. However, $q_g>10^{-9} s^{-2}$ lies
outside our parameter space of interest. At lower $q_g$ we were unable
to construct models at the threshold, and simply extrapolated this
fit. $q_g=10^{-12} s^{-2}$ at 160 $R_g$ for a 10$^8
\Msun$ BH.  Note that the spectrum in the right-hand panel of Fig.~\ref{fig:composite} 
would not satisfy our conservative criterion. We also impose a low-temperature threshold of
10$^4$ K, below which metal absorption becomes important and possibly
causes metal edges to appear in the spectrum.  Comparing the maximum
disc temperature to the fitting formula above, we identify regions in
binary parameter space which are cool enough to have Lyman
edges.\footnote{Using the hottest annulus, as opposed to full composite spectra, is a good proxy
  to identify the presence/absence of the Lyman edge.}  In
Fig.~\ref{fig:param_space}, we show the ranges of masses and
separations for which a prominent Lyman edge is produced, for two
different accretion rates and mass ratios. The corresonding ranges of orbital periods, and time-scale 
over which the edge would be visible, may be calculated (to within a factor of a few) by the following fitting formulas. 

\begin{align}
 t_{\rm orb} &\simeq 0.04 - 0.09 \; {\rm yr} \left(\frac{\eddr}{0.1}\right)^{1/4} \left(\frac{M}{10^7 \Msun}\right)^{1/3} \left(\frac{q}{1000}\right)^{1/10}\\
 t_{\rm ly} &\simeq 10^4 \; {\rm yr} \left(\frac{\eddr}{0.1}\right)^{-1/2} \left(\frac{M}{10^7 \Msun}\right)^{1/3} \left(\frac{q}{1000}\right)^{3/4}
 \end{align}

Note that for a given $\eddr$ and $q$, there is a maximum total mass for the edge which is set by our low temperature threshold:

\begin{align}
 \left(\frac{M}{10^7 \Msun}\right) \lesssim 50 \left(\frac{\eddr}{0.1}\right)^{3/5} \left(\frac{q}{1000}\right)^{-3/5}
\end{align}

\section{Mini-discs}
\label{sec:minidiscs}

In general, we would not expect any central cavity to be completely
empty. As first discussed in the smoothed-particle hydrodynamic simulations of
\citet{ArtyLubow:1996} and confirmed by several recent works
\citep{Hayasaki:2007,MacFadyenMilos2008,Cuadra:2009,ShiKrolik:2012,Roedig:2012,Dorazio+2013,Farris+2013,Gold+2014}
gas leaks into the cavity through non-axisymmetric streams. These
streams can feed ``mini-discs'' around each individual BH, at rates
set by the viscous time-scale of each mini-disc. The recent simulations
suggest that accretion rates inside the cavity may be comparable to
those onto a single-BH.

\begin{figure}
 \centering
 \mbox{\includegraphics[width=0.48\linewidth]{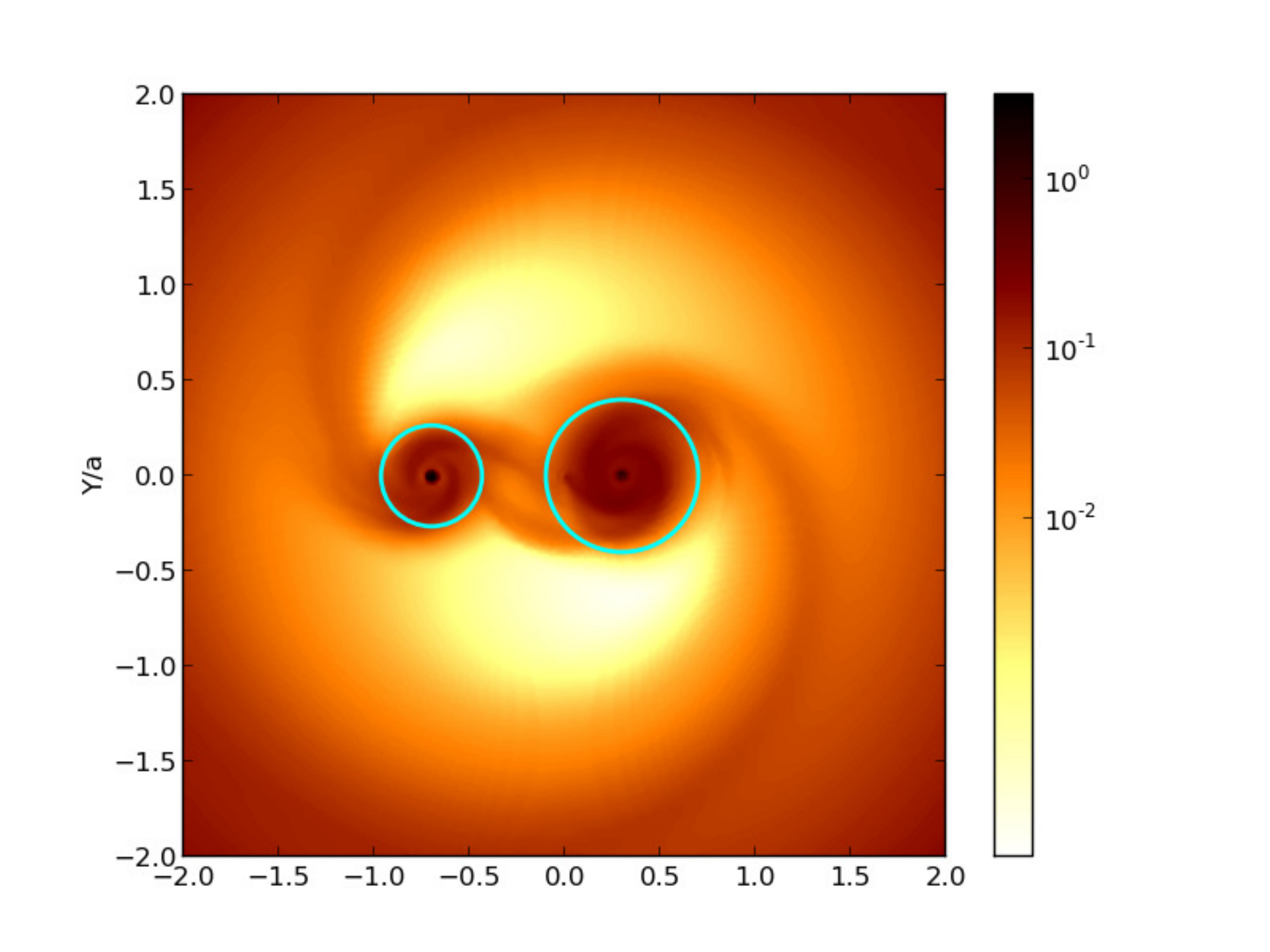}}
 \mbox{\includegraphics[width=0.48\linewidth]{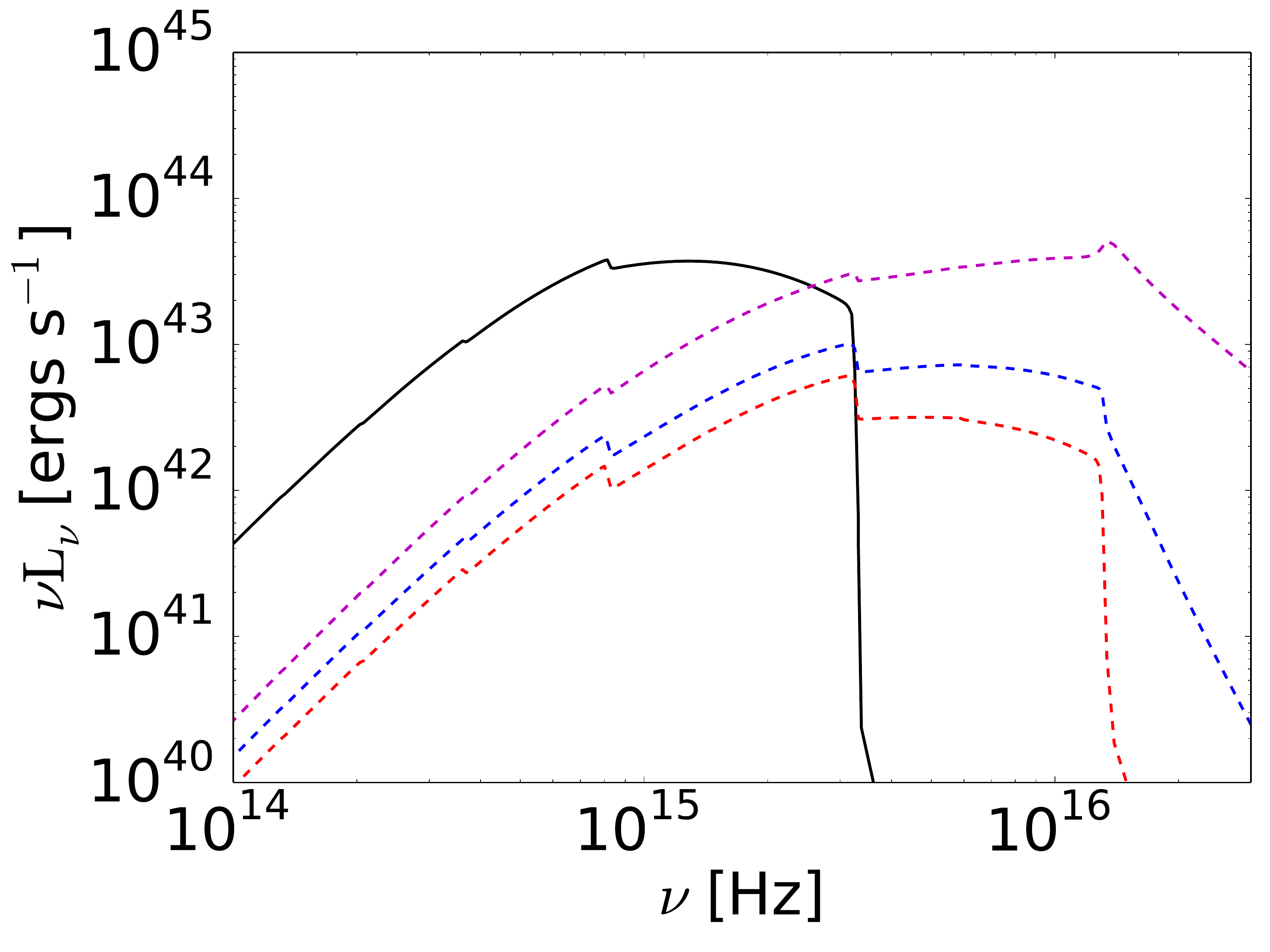}}
 \caption{{\em Left:} Time average of the surface density distribution,
   reproduced with permission from
   \citet[][Fig. 8b]{Farris+2013}. {\em Right:} spectra for the
   circumbinary and circumsecondary discs. The solid (black) curve is
   for the circumbinary disc, and the other curves are for the
   mini-disc, corresponding to a different assumed fraction of the
   circumbinary $\dot{M}$ fueling the secondary BH: 50\% (purple), 10\% (blue), and 5\% (red).}
 \label{fig:minidiscs}
\end{figure}

The emission from hot mini-discs may mask the Lyman edge feature. To
illustrate this, we calculate spectra for the discs shown in Fig. 8b
of \citealt{Farris+2013} (reproduced here as the left panel of
Fig.~\ref{fig:minidiscs}).  This figure shows a time averaged
surface density profile for a $q=0.43$ binary. The simulation is scale-free so
the masses, separation, and accretion rate are arbitrary.  We adopt a
$M_{\rm tot}=5 \times 10^7 \Msun$ and separation $r_s=460 R_g$.  For $q=0.43$, the masses of the primary and secondary are then
$3.5\times10^7 \; \Msun$ and $1.5 \times 10^7 \; \Msun$, and the
system is on the verge of the GW inspiral stage.

To model the spectrum of the circumbinary disc, we use $\teff(r)$ from
an axisymmetric KHL12 circumbinary disc model with inner radius at 460
$R_g$ (even though the simulated disc is in fact lopsided; note that
the simulations are isothermal and do not predict $\teff$). To obtain
the $\teff$ profile, we set $\dot{M}/\dot{M}_{\rm Edd}=0.1$.  We
likewise model the circumsecondary disc as a standard thin disc
around a non-spinning BH (with truncation radius of 400 $R_g$, where $R_g$ is defined in terms
of just the secondary mass. This is roughly consistent
with the size of the circumsecondary disc in Fig.~\ref{fig:minidiscs}).  
Fig.~\ref{fig:minidiscs} shows spectra of the circumbinary and circumsecondary discs, assuming that 50\% (purple),
10\% (blue), and 5\% (red) of the external $\dot{M}$ fuels the
secondary. For the 5\% case, the Lyman edge is still
prominent. However, it is greatly reduced for 10\%, and for 50\% it is
completely obscured. Thus, if more than a few \% of the $\dot{M}$ in
the cicumbinary disc leaks into the cavity and fuels a radiatively
efficient, hot accretion flow, the Lyman edge can be obscured. As discussed
above, \citet{Farris+2013} find circumsecondary accretion rates
comparable to the rate onto a single BH, which would favor the $>50\%$
case. There should also be some emission from the circumprimary disc,
which we have not included.

However, there are two reasons why Lyman edges could remain
detectable, even with efficient fueling of the
individual BHs.  First, most of the gas entering the cavity may fuel
the secondary (rather than the primary) BH. This could then lead to a
radiatively inefficient, super-Eddington accretion flow, rather than a
thin mini-disc; such discs have much fainter fluxes in the UV (see,
e.g. \citealt{Kawaguchi2003}).  Second, the fueling of the individual
BHs may be intermittent.  The simulations listed in the preceding
paragaphs show that the rate at which the gas enters the cavity
fluctuates strongly, tracking the binary's orbital period.  Whether
the BHs can accept this fuel depends on the viscous time-scale in their
vicinity.  There is some evidence from 3D magnetohdydrodynamic simulations that the
effective $\alpha$ may strongly increase inside the cavity
\citep{ShiKrolik:2012,Noble+2012,Gold+2014}. The streams from the
circumbinary disc would then rapidly accrete onto the individual BHs,
and mini-discs would either not form or would be intermittent
\citep{Tanaka:2013}. The relevant time-scale would be the viscous time
of the mini-disc, which should generally extend to the tidal truncation radius
as long as the specific angular momentum of the accreting streams exceeds that at the ISCO \citep{Roedig+:2014}. 
Ultimately, the prominence of the Lyman edge
features is tied to the nature of the mini-disks, and requires a better understanding
of these flows.

\section{Summary and Conclusions} 
\label{sec:conclusion}

In this Letter, we have proposed that spectral edges, in
particular at the Lyman limit, may be characteristic signatures of a
circumbinary disc. Our conclusions can be summarized as follows.

\begin{enumerate}

 \item If binary torques clear a cavity in the circumbinary disc, the
   disc spectrum may exhibit a sharp drop at the Lyman limit. This is
   because the hottest region (i.e. the inner edge) of the disc is
   cool enough to have neutral H, absorbing nearly all flux blueward
   of the Lyman limit. This occurs below a critical $\teff$ which
   generally lies in the range $\sim$10,000 K-20,000 K, depending on
   vertical gravity and other parameters.  At lower temperatures,
   absorption from metals (i.e. C) may cause spectral edges redward of
   the Lyman limit.

 \item Observationally, AGN spectra only show Lyman edges due to
   absorption by intervening neutral gas (see \citealt{Antonucci+:1989}). The inner regions of a
   single-BH AGN disc are hotter than for binaries
   (Fig.~\ref{fig:temperature}), and can mask any edge produced in the
   outer disc, leaving only a small
   ``kink''(Fig.~\ref{fig:composite}).  Such kinks are not seen
   observationally, and understanding
   what would smear them (e.g. general relativistic effects or winds)
   is an open theoretical problem. For an overview of the
   Lyman edge problem in AGN spectral modeling see,
   e.g. \citet{KolykhalovSunyaev1984} and \citet{KoratkarBlaes:99}.

 \item Neutral H in the binary's host galaxy, unrelated to the nuclear
   accretion disc itself, could cause a Lyman edge (as seen in a few
   AGNs).  However, in the case of the disc, one could look for
   rotational broadening of the edge due to orbital motions in the
   disc, with velocities of order $10^4$ km/s.  This may cause a
   $\simeq$10\% smearing on the edge.

 \item A portion of the binary parameter space could have
   a truncated circumbinary disc, with $\teff$ in the critical range
   for a prominent Lyman edge (Fig.~\ref{fig:param_space}). This
   parameter space partially overlaps with the expected typical
   parameters for individually resolvable PTA sources. These are very
   massive binaries, with $10^8 \Msun \lsim M \lsim 10^9 \Msun$, and
   separations ranging from 10's to 1000's of $R_g$, and mass ratios
   peaking at $q\sim1$ but with a long tail to lower values
   \citep{Sesana+2012}.

 \item Efficient fueling of the BHs inside the central cavity could
   mask the Lyman edge feature in the circumbinary disc spectrum. For
   example, persistent emission from hot mini-discs
   (see~\citealt{Farris+2013}) would obscure the Lyman edge.  However,
   if the accretion flows onto the individual BHs are radiatively
   inefficient and/or intermittent \citep{Tanaka:2013}, the Lyman
   edge could remain visible, or appear periodically on the time-scale
   of the binary's orbit (which could be weeks to years; \citealt{HKM09}).
   
 \item The proposed Lyman edge signature could be used in combination with other proposed EM signatures
 to refine the search for SMBHBs. We have conducted a preliminary search for the Lyman edge feature 
 in x-ray weak quasars discussed in \citealt{Brandt+:2000}. In particular we looked at FUSE and IUE spectra for 
 the ten objects in their Table 2, but found no sign of any Lyman edge feature. 
 
\end{enumerate}

If detected in an AGN, a prominent Lyman edge would tighten the case
for the presence of a compact binary BH.

\vspace{\baselineskip} We thank Shane Davis for sharing a modified
version of \texttt{TLUSTY} and for technical help, as well as for
providing an initial table of spectral models. We also thank Shane
Davis, Omer Blaes, Ivan Hubeny, Jules Halpern, and Frits Paerls for
insightful conversations.  ZH acknowledges support from NASA grant
NNX11AE05G.

\footnotesize{
  \bibliographystyle{mn2e}
  \bibliography{master}
}
\end{document}